\documentclass[%ｎ%runinaddress,
showpacs,
 amsmath,amssymb,
prb,
twocolumn,
]{revtex4-2}
\usepackage{graphicx}% Include figure files
\usepackage{dcolumn}% Align table columns on decimal point
\usepackage{bm}% bold math
\usepackage{braket}
\usepackage{color}
\usepackage{accents}
\usepackage{mathrsfs}
\usepackage{bm}
\usepackage{here}
\usepackage{longtable}

\begin{document}
%\preprint{APS/123-QED}
%%%%%%%%%%%%%%%
\title{
Spin Susceptibility of a $J=3/2$ Superconductor}
\author{Dakyeong Kim$^{1}$}
\author{Takumi Sato$^{1}$}
\author{Shingo Kobayashi$^{2}$}
\author{Yasuhiro Asano$^{1}$}%
\affiliation{
$^{1}$Department of Applied Physics, Hokkaido University, Sapporo 060-8628, Japan.\\
$^{2}$RIKEN Center for Emergent Matter Science, Wako, Saitama 351-0198, Japan.
}
%%%%%%%%%%%%%%%
\date{\today}
%%%%%%%%%%%%%%%
\begin{abstract}
We discuss the spin susceptibility of superconductors in which a Cooper pair consists of two electrons 
having the angular momentum $J=3/2$ due to strong spin-orbit interactions.
The susceptibility is calculated analytically for several pseudospin quintet states in a cubic superconductor 
within the linear response to a Zeeman field.
The susceptibility for $A_{1g}$ symmetry states is isotropic in real space.
For $E_g$ and $T_{2g}$ symmetry cases, the results depend sensitively on choices of order parameter.
The susceptibility is isotropic for a $T_{2g}$ symmetry state, whereas it becomes anisotropic 
for an $E_{g}$ symmetry state. We also find in a $T_{2g}$ state that the susceptibility tensor
has off-diagonal elements. 
\end{abstract}
%%%%%%%%%%%%%%%
\maketitle
%%%%%%%%%%%%%%%

%%%%%%%%%%%%%%%%%%%%%%%%%%%%%%%%%%%%%%%%%%%%
\section{Introduction}
%%%%%%%%%%%%%%%%%%%%%%%%%%%%%%%%%%%%%%%%%%%%
Spin-orbit interaction is a source of exotic electronic states 
realized in topological semimetals~\cite{chen:cpb2016,armitage:rmp2018}, topological insulators~\cite{hasan:rmp2010,qi:rmp2011}, and topological superconductors~\cite{tanaka:jpsj2011,sato:jpsj2016,mizushima:jpsj2016,chiu:rmp2016,sato:rpp2017,sato:rpp2017}.
In the presence of strong spin-orbit interactions, 
spin $S=1/2$ and orbital angular momentum $L=1$ of an electron are inseparable degrees of freedom.
Electronic properties of such materials are characterized by an electron 
with pseudospin $J=L+S=3/2$.
Recent studies have suggested a possibility of superconductivity 
due to Cooper pairing between two electrons with $J=3/2$~\cite{butch:prb2011,nakajima:sciadv2015}.
The large angular momentum of an electron enriches the symmetry of the order parameter such as
pseudospin-quintet even-parity and pseudospin-septet odd-parity~\cite{brydon:prl2016,agterberg:prl2017}
in addition to conventional spin-singlet even-parity and spin-triplet odd-parity.
Such high angular-momentum pairing states would feature superconducting phenomena 
of $J=3/2$ superconductors~\cite{brydon:prb2018,kim:sciadv2018,venderbos:prx2018,kobayashi:prl2019,kim:prb2021,kim:jpsj2021}. 
In particular, a large angular momentum of a Cooper pair would qualitatively change 
the magnetic response of a superconductor to an external magnetic field.

 The spin susceptibility reflects well the internal spin structures of a Cooper pair.
%When a temperature is higher than the transition temperature $T_c$, 
%the magnetic response of normal metals is paramagnetic and the 
%spin susceptibility is independent of temperatures.
It is well known in spin-singlet superconductors that the
 spin susceptibility decreases monotonically with the decrease of temperature below 
 $T_c$ and vanishes at zero temperature~\cite{yoshida:pr1958}.
This phenomenon occurs independently of the direction of a Zeeman field $\boldsymbol{H}$ 
because a Cooper pair has no spin. In spin-triplet superconductors, on the other hand, 
the susceptibility can be anisotropic depending on the relative 
alignment between a Zeeman field and a $\boldsymbol{d}$ vector in the order parameter. 
For $\boldsymbol{d} \perp \boldsymbol{H}$, the spin susceptibility is constant independent of 
temperature. Thus, the unchanged Knight shift across $T_c$ in experiments 
could be a strong evidence of spin-triplet superconductivity.
For $\boldsymbol{d} \parallel \boldsymbol{H}$, the susceptibility decreases with decreasing temperature below $T_c$.
Such anisotropy is more remarkable when the number of components in a $\boldsymbol{d}$ vector is smaller.
For $J=3/2$ superconductors, however, our knowledge of the spin susceptibility is very limited to 
a theoretical paper that reported vanishing the spin susceptibility at zero temperature 
for a singlet-quintet mixed state in a centrosymmetric superconductor~\cite{yu:jap2020}.

In this paper, we study theoretically the response of 
 pseudospin-quintet even-parity superconductors to an external Zeeman field.
The angular momentum of a Cooper pair in such superconductors is $J=2$.
Since the pairing symmetries of the pseudospin-quintet states are not well understood, 
we decided to calculate the spin susceptibility for the plausible pair 
potentials at a cubic superconductor preserving time-reversal symmetry.
The spin susceptibility is analytically calculated based on the linear 
response formula~\cite{mahan}. 
The pair potential of pseudospin-quintet states
is described by a five-component vector that couples to five $4 \times 4$ 
matrices in pseudospin space. 
Such complicated internal structures of the pair potential enrich the magnetic response of 
$J=3/2$ superconductors. 
When the symmetry of the pair potential is high in pseudospin space ($A_{1g}$ state), 
the magnetic response is isotropic in real space and the spin susceptibility decreases monotonically 
with the decrease of temperature. 
The results are similar to those of $^3$He B-phase.
When the pair potential is independent of wavenumber ($T_{2g}$ and $E_g$ states), 
the spin susceptibility shows unique features to pseudospin-quintet states.
The magnetic response becomes anisotropic in real space in an $E_g$ state 
and the susceptibility tensor has finite off-diagonal elements in a $T_{2g}$ state.

This paper is organized as follows.
In Sec.~II, we described the electronic structure and the pair potential of a
$J=3/2$ superconductor in terms of five $4\times 4$ matrices. 
The characteristic behaviors of the spin-susceptibility are discussed 
in Sec.~III.
The conclusion is given in Sec.IV. Algebras of $4 \times 4$ matrices and 
a number of  
mathematical relationships used in the paper are summarized in Appendices.
Throughout this paper, we use the system of units $\hbar=k_B=c=1$, where $k_B$ is the Boltzmann constant 
and $c$ is the speed of light.

%%%%%%%%%%%%%%%%%%%%%%%%%%%%%%%%%%%%%%%%%%%%
\section{J=3/2 Superconductor}
%%%%%%%%%%%%%%%%%%%%%%%%%%%%%%%%%%%%%%%%%%%%

 We begin our analysis with the normal state Hamiltonian adopted 
in Ref.~\onlinecite{agterberg:prl2017}.
The electronic states have four degrees of freedom consisting of two orbitals of equal parity and spin 1/2. 
In the presence of strong spin-orbit interactions, the effective Hamiltonian for a $J=3/2$ electron
is given by
~\cite{luttinger:pr1955}
\begin{align}
\mathcal{H}_{\mathrm{N}} =& \sum_{\boldsymbol{k}} 
\Psi_{\boldsymbol{k}}^\dagger \, H_{\mathrm{N}}(\boldsymbol{k}) \, \Psi_{\boldsymbol{k}},\\ 
\Psi_{\boldsymbol{k}} =& \left[ c_{\boldsymbol{k}, 3/2}, 
c_{\boldsymbol{k}, 1/2}, c_{\boldsymbol{k}, -1/2}, c_{\boldsymbol{k}, -3/2}\right]^{\mathrm{T}},
\end{align}
where $\mathrm{T}$ means the transpose of a matrix and 
$c_{\boldsymbol{k}, j_z}$ is the annihilation operator of an electron at $\boldsymbol{k}$ with 
the $z$-component of angular momentum being $j_z$.
The normal state Hamiltonian is represented by
\begin{align}
H_{\mathrm{N}}(\boldsymbol{k}) =& \alpha \boldsymbol{k}^2 + \beta 
\left(\boldsymbol{k}\cdot \boldsymbol{J}\right)^2 - \mu=
 \xi_{\boldsymbol{k}} \, 1_{4\times 4}
+ \vec{\epsilon}_{\boldsymbol{k}} \cdot \vec{\gamma}
 \label{eq:hn_cubic}
\end{align}
with $\xi_{\boldsymbol{k}}= \epsilon_{\boldsymbol{k},0} -\mu$ and
\begin{align} 
\epsilon_{\boldsymbol{k}, 0} = &\left(\alpha+ \frac{5}{4}\beta\right) \boldsymbol{k}^2, \quad
\epsilon_{\boldsymbol{k},j} = \beta\, \boldsymbol{k}^2 \, e_j\, c_j(\hat{\boldsymbol{k}}), \label{eq:ep_def}\\
c_1(\hat{\boldsymbol{k}}) =& \sqrt{15}\,  \hat{k}_x\, \hat{k}_y, \quad   
c_2(\hat{\boldsymbol{k}}) = \sqrt{15}\,  \hat{k}_y\, \hat{k}_z, \\
c_3(\hat{\boldsymbol{k}}) =& \sqrt{15}\,  \hat{k}_z\, \hat{k}_x, \quad  
c_4(\hat{\boldsymbol{k}}) =  \frac{\sqrt{15}}{2}\,   (\hat{k}_x^2- \hat{k}_y^2), \\ 
c_5(\hat{\boldsymbol{k}}) =&  \frac{\sqrt{5}}{2}\,  (2\hat{k}_z^2 - \hat{k}_x^2- \hat{k}_y^2),
\end{align}
where $\hat{k}_j= k_j/|\boldsymbol{k}|$ for $j=x, y$ and $z$ 
represents the direction of wavenumber on the Fermi surface.
The constants $\alpha>0$ and $\beta$ determine the normal state property.
%The normal state are degenerate for both spins and orbitals at $\beta=0$.
 The spin-orbit interactions increase with the increase of $\beta>0$.
%As a result, the dispersion depends on pseudospin.
%The unit vector $\vec{e}$ determines the electronic structure in pseudospin space.
The coefficients $c_j$ are normalized as
\begin{align}
\langle c_i(\hat{\boldsymbol{k}}) \; c_j(\hat{\boldsymbol{k}}) \rangle_{\hat{\boldsymbol{k}}} \equiv
\int \frac{d \hat{\boldsymbol{k}}} {4\pi} \,c_i(\hat{\boldsymbol{k}})  c_j(\hat{\boldsymbol{k}}) =\delta_{i, j}, \label{eq:orthonoaml_cij}
\end{align}
where $\langle \cdots \rangle_{\hat{\boldsymbol{k}}} $ means the integral over the solid angle on the Fermi surface.
The normalized five-component vector $\vec{e}=(e_1, e_2, e_3, e_4, e_5)/|\vec{e}|$ determines 
the dependence of the normal state dispersions on pseudospins.
%It is usually chosen as $\vec{e}=(1, 1, 1, 1, 1)/\sqrt{5}$~\cite{luttinger:pr1955}.
%$\vec{\epsilon}_{\boldsymbol{k}}$ represents a five-component vector and 
The spinors for the angular momenum of $J=3/2$ are described by, 
\begin{align}
J_x =& \frac{1}{2}\left[\begin{array}{cccc}
0 & \sqrt{3} & 0 & 0 \\
\sqrt{3} & 0 & 2 & 0\\
0 & 2 & 0 & \sqrt{3} \\
0 & 0 & \sqrt{3} & 0
\end{array}\right], \\
J_y =& \frac{1}{2}\left[\begin{array}{cccc}
0 & -i\sqrt{3} & 0 & 0 \\
i\sqrt{3} & 0 & -2i & 0\\
0 & 2i & 0 & -i\sqrt{3} \\
0 & 0 & i\sqrt{3} & 0
\end{array}\right], \\
J_z =& \frac{1}{2}\left[\begin{array}{cccc}
3 & 0 & 0 & 0 \\
0 & 1 & 0 & 0\\
0 & 0 & -1 & 0 \\
0 & 0 & 0 & -3
\end{array}\right].
\end{align}
The $4 \times 4$ matrices in pseudospin space are defined as
\begin{align}
\gamma^1=& \frac{1}{\sqrt{3}} (J_x J_y +J_y J_x), \quad
\gamma^2= \frac{1}{\sqrt{3}} (J_y J_z +J_z J_y), \label{eq:gamma1}\\
\gamma^3=& \frac{1}{\sqrt{3}} (J_z J_x +J_x J_z), \quad
\gamma^4= \frac{1}{\sqrt{3}} (J_x^2- J_y^2), \\
\gamma^5=& \frac{1}{3} (2J_z^2 - J_x^2 -J_y^2), \label{eq:gamma5}
\end{align}
and $1_{4\times 4}$ is the identity matrix.
They satisfy the following relations
\begin{align}
&\gamma^\nu\, \gamma^\lambda + \gamma^\lambda\, \gamma^\nu =2 \times  1_{4\times 4} \delta_{\nu, \lambda}, \\
&\gamma^1\, \gamma^2\, \gamma^3\, \gamma^4\, \gamma^5=-1_{4\times 4},\\ 
&\{\gamma^\nu\}^\ast = \{\gamma^\nu\}^{\mathrm{T}}= U_T\, \gamma^\nu \, U_T^{-1},\quad U_T=\gamma^1\, \gamma^2,
\end{align}
where $U_T$ is the unitary part of the 
time-reversal operation $\mathcal{T}=U_T \, \mathcal{K}$ with $\mathcal{K}$ meaning complex conjugation.
The superconducting pair potential is represented as
\begin{align}
\Delta(\boldsymbol{k}) = \vec{\eta}_{\boldsymbol{k}} \cdot \vec{\gamma} \, U_T, \label{eq:pairpotential0}
\end{align}
where a five-component vector $\vec{\eta}_{\boldsymbol{k}}$ represents
an even-parity pseudospin-quintet state. 
Throughout this paper, we assume that all components of $\vec{\eta}_{\boldsymbol{k}}$ are real values.
As a result of the Fermi-Dirac statistics of electrons, the pair potential is antisymmetric 
under the permutation of two pseudospins, (i.e., 
 $\Delta^{\mathrm{T}}(\boldsymbol{k}) = - \Delta(\boldsymbol{k}) $). 
The Bogoliubov-de Gennes 
Hamiltonian reads,
\begin{align}
H_{\mathrm{BdG}}(\boldsymbol{k})= \left[\begin{array}{cc}
H_{\mathrm{N}}(\boldsymbol{k}) & \Delta(\boldsymbol{k}) \\
- \undertilde{\Delta}(\boldsymbol{k})  & -\undertilde{H}_{\mathrm{N}}(\boldsymbol{k})
\end{array}\right],
\end{align}
where $\undertilde{X}(\boldsymbol{k}, i\omega) \equiv X^\ast(-\boldsymbol{k}, i\omega)$ represents 
the particle-hole conjugation of $X(\boldsymbol{k}, i\omega)$.

%%%%%%%%%%%%%%%%%%%%%%%%%%%%%%%%%%%%%%%%%%%%
%\section{Formula }
%%%%%%%%%%%%%%%%%%%%%%%%%%%%%%%%%%%%%%%%%%%%
The interaction with a uniform Zeeman field $\boldsymbol{H}$ is described by~\cite{luttinger:pr1955}
\begin{align}
H_{\mathrm{Z}}=& - \mu_B  \tilde{\boldsymbol{J}} \cdot \boldsymbol{H}, \\
\tilde{J}_j =&   g_1 {J}_j + g_{3} {J}_j^3, 
\end{align}
for $j=x, y,$ and $z$, 
where $\mu_B$ is the Bohr's magneton, and $g_1$ and $g_3$ are the coupling constants. 
The matrix structures of $J_j$ and $J_j^3$ are displayed in Appendix~\ref{ap:normal}. 
The angular momenta in the Zeeman Hamiltonian are then given by
\begin{align}
\tilde{J}_x =& -\frac{i}{2} \left( 
\sqrt{3} p_1   \gamma^2\, \gamma^5 + p_2  \gamma^1\, \gamma^3 + p_1 \gamma^2\, \gamma^4 \right), \\
\tilde{J}_y =& \frac{i}{2} \left( 
\sqrt{3} p_1   \gamma^3\, \gamma^5 + p_2  \gamma^1\, \gamma^2 - p_1 \gamma^3\, \gamma^4 \right), \\
\tilde{J}_z =& \frac{i}{2} \left( 
 p_2   \gamma^2\, \gamma^3 + 2 p_1  \gamma^1\, \gamma^4 \right), \\
p_1 =& g_1 + \frac{7}{4} g_3, \quad p_2 = g_1 + \frac{13}{4} g_3.
\end{align}
%
%Therefore, the $J_j^3$ term in the Zeeman Hamiltonian does not change qualitatively the
%magnetic response due to $J_j$ term.
%In this paper, we neglect the 
%effects of $\boldsymbol{J}^3$ by choosing $g^\prime=0$.
In the linear response theory, the spin susceptibility is calculated by using 
the formula~\cite{mahan}
\begin{align}
\chi_{\mu \nu}
= &\chi_{\mathrm{N}} \, \delta_{\mu, \nu} 
-
\left(\frac{\mu_B}{2}\right)^2 
 T\sum_{\omega_n} 
\int\frac{d \boldsymbol{k}}{(2\pi)^3} \nonumber\\
\times  \mathrm{Tr} &\left[
{G}(\boldsymbol{k}, i\omega_n) \, \tilde{J}_\mu 
\,
G(\boldsymbol{k}, i\omega_n) \, \tilde{J}_\nu 
\right. \nonumber\\
&
+
\undertilde{F}(\boldsymbol{k}, i\omega_l)\, \tilde{J}_\mu \,
{F}(\boldsymbol{k}, i\omega_n) \, (\tilde{J}_\nu )^\ast \nonumber\\
&\left.-{G}_{\mathrm{N}}(\boldsymbol{k}, i\omega_n) \, \tilde{J}_\mu 
\,
G_{\mathrm{N}}(\boldsymbol{k}, i\omega_n) \, \tilde{J}_\nu
\right]. \label{eq:regular_int}
\end{align}
The summation over the Matsubara frequency and that over the wavenumber are regularized 
by introducing the Green's functions in the normal state ${G}_{\mathrm{N}}$ and 
the spin susceptibility $\chi_{\mathrm{N}}$ in the normal state~\cite{agd}.

 The Green's function for a superconducting state can be 
obtained by solving the Gor'kov equation
\begin{align}
&\left[\begin{array}{cc}
i\omega_n - H_{\mathrm{BdG}}(\boldsymbol{k}) 
\end{array}\right]
\left[\begin{array}{cc} G(\boldsymbol{k}, i\omega_n) & F(\boldsymbol{k}, i\omega_n) \\
 -\undertilde{F}(\boldsymbol{k}, i\omega_n) & -\undertilde{G}(\boldsymbol{k}, i\omega_n)
\end{array}\right] \nonumber\\
 &= 1_{8 \times 8}. \label{eq:gorkov}
\end{align}
The anomalous Green's function results in 
\begin{align}
%G^{-1}&(\boldsymbol{k}, i\omega_n) 
%=& \left[ i\omega_n - H_N + \Delta\, (i\omega_n + \undertilde{H}_N)^{-1} \,
%\undertilde{\Delta} \right],\\
%= i\omega_n - \xi_{\boldsymbol{k}}  - \vec{\epsilon}_{\boldsymbol{k}}\cdot \vec{\gamma} \nonumber\\
%&- 
%\vec{\eta}_{\boldsymbol{k}}\cdot \vec{\gamma} 
%\frac{i\omega_n + \xi_{\boldsymbol{k}}  - \vec{\epsilon}_{\boldsymbol{k}}\cdot \vec{\gamma}}
%{(i\omega_n + \xi_{\boldsymbol{k}} )^2- {\vec{\epsilon}_{\boldsymbol{k}}}^2}\, 
%\vec{\eta}_{\boldsymbol{k}}\cdot \vec{\gamma},\\ 
% 
F^{-1}&(\boldsymbol{k}, i\omega_n)
%=& 
%\left[ 
%\undertilde{\Delta} + (i\omega_n + \undertilde{H}_N)  \,  \Delta^{-1}\, 
%(i\omega_n - H_N) \right], \\
%=& -U_T\left[
%- \boldsymbol{\eta}\cdot \boldsymbol{\gamma} +
%(i\omega_n + \xi + \boldsymbol{\epsilon}\cdot \boldsymbol{\gamma}) 
%\frac{\boldsymbol{\eta}\cdot \boldsymbol{\gamma} }{\boldsymbol{\eta}^2}
%(i\omega_n - \xi -\boldsymbol{\epsilon}\cdot \boldsymbol{\gamma})\right],\\
=\frac{U_T}{{\vec{\eta}_{\boldsymbol{k}}}^2}
\left[
(\omega_n^2+{\xi_{\boldsymbol{k}} }^2+ {\vec{\eta}_{\boldsymbol{k}}}^2) \vec{\eta}_{\boldsymbol{k}} \cdot \vec{\gamma}
\right. \nonumber\\
&+ i\omega_n [\vec{\eta}_{\boldsymbol{k}} \cdot \vec{\gamma}, \vec{\epsilon}_{\boldsymbol{k}}
\cdot \vec{\gamma}]_{-}
 \nonumber\\
&\left.
+ 2\, \xi_{\boldsymbol{k}}  \,\vec{\eta}_{\boldsymbol{k}} \cdot\vec{\epsilon }_{\boldsymbol{k}}
+ (\vec{\epsilon}_{\boldsymbol{k}}\cdot \vec{\gamma}) 
\, (\vec{\eta}_{\boldsymbol{k}}\cdot \vec{\gamma}) \, 
(\vec{\epsilon}_{\boldsymbol{k}} \cdot \vec{\gamma})
\right]. \label{eq:fs_general}
\end{align}
Generally speaking, it is not easy to calculate analytically the inversion of $4 \times 4$ matrices.

To proceed analytic calculation, we consider a cubic symmetric superconductor~\cite{brydon:prb2018}, in which even-parity pair potentials are classified into $A_{1g}$, $E_g$, and $T_{2g}$ states according to irreducible representations (irreps) of cubic symmetry. With focusing on pseudospin-quintet Cooper pairs, their pairing states are explicitly represented as
\begin{align}
A_{1g}: \quad &\vec{\eta}_{\boldsymbol{k}} \cdot \vec{\gamma}  = 
\Delta \, \sum_{j=1}^5 h_j\, c_j(\hat{\boldsymbol{k}})\, \gamma^j, \label{eq:eta_d} \\
E_{g}: \quad &\vec{\eta}_{\boldsymbol{k}} \cdot \vec{\gamma}  = 
\Delta \, ( l_4 \gamma^4+ l_5 \gamma^5), \label{eq:eta_eg} \\
T_{2g}: \quad &\vec{\eta}_{\boldsymbol{k}} \cdot \vec{\gamma}  = 
\Delta \, ( l_1 \gamma^1+ l_2 \gamma^2 +  l_3 \gamma^3), \label{eq:eta_t2g} 
\end{align}
where the unit vector $\vec{h}$ determines the pseudospin structure of $A_{1g}$ state and $l_i \in \mathbb{C}$ ($i=1-5$).  $A_{1g}$ states involve momentum-dependent coefficients, which comes from the fact that $\gamma^4$ and $\gamma^5$ $(\gamma^1,\gamma^2$, and $\gamma^3)$ themselves belong to $E_g$ ($T_{2g}$) irreps of cubic symmetry.

Generally speaking, the coefficients $l_i$ for $i=1-5$ are 
determined from the steady states of the free energy~\cite{sigrist:rmp1991}. 
For $E_g$ state, three distinct steady states exist: $(l_4,l_5)=(1,0)$, $(0,1)$, and $(1,i)/\sqrt{2}$. 
The first two states preserve time-reversal symmetry, while the last state breaks time-reversal symmetry. 
For $T_{2g}$ state, there are four distinct steady states: $(l_1,l_2,l_3)=(1,1,1)/\sqrt{3}$, $(1,0,0)$, $(1,e^{i 2\pi/3},e^{i 4\pi/3})/\sqrt{3}$, $(1,i,0)/\sqrt{2}$. The last two states break time-reversal symmetry.
Time-reversal symmetry-breaking superconducting states have a problem specific to 
them: the formation of Bogoliubov-Fermi surfaces~\cite{agterberg:prl2017,brydon:prb2018,kim:jpsj2021}.
Since the relations between the pseudospin structures of a Cooper pair and the magnetic
response of a superconductor is the main issue in this paper,
we focus on time-reversal symmetry respecting superconducting states.
%In the following, we focus on time-reversal symmetric states, where $l_i$ ($i=1-5$) take a real number. 
In addition to the steady states, 
we also consider another pseudospin states described by 
$(l_4, l_5) =(1,1)/\sqrt{2}$, $(l_1, l_2,l_3) =(1,1,0)/\sqrt{2}$, and $(l_1,l_2,l_3,l_4,l_5) =(1,1,1,1,1)/\sqrt{5}$.
The last one is the admixture of $E_{g}$ and $T_{2g}$ states. 
These states complement possible combinations of $\gamma^i$ ($i=1-5$).
The comparison between the calculated results for such states and those for the steady states 
helps us to understand the relations between the pseudospin structures of a Cooper pair and 
the spin susceptibility.
Note that $(l_1, l_2,l_3) =(1,1,0)/\sqrt{2}$ represents a possible order parameter 
in tetragonal symmetric superconductors with the high symmetry axis being the $y$ direction~\cite{sigrist:rmp1991}.

% =========================================
\section{Spin susceptibility}
% =========================================
%-----------------------------------------------------
%----------------------------
%\onecolumngrid
\begin{figure}[tbp]
\begin{center}
  \includegraphics[width=9.0cm]{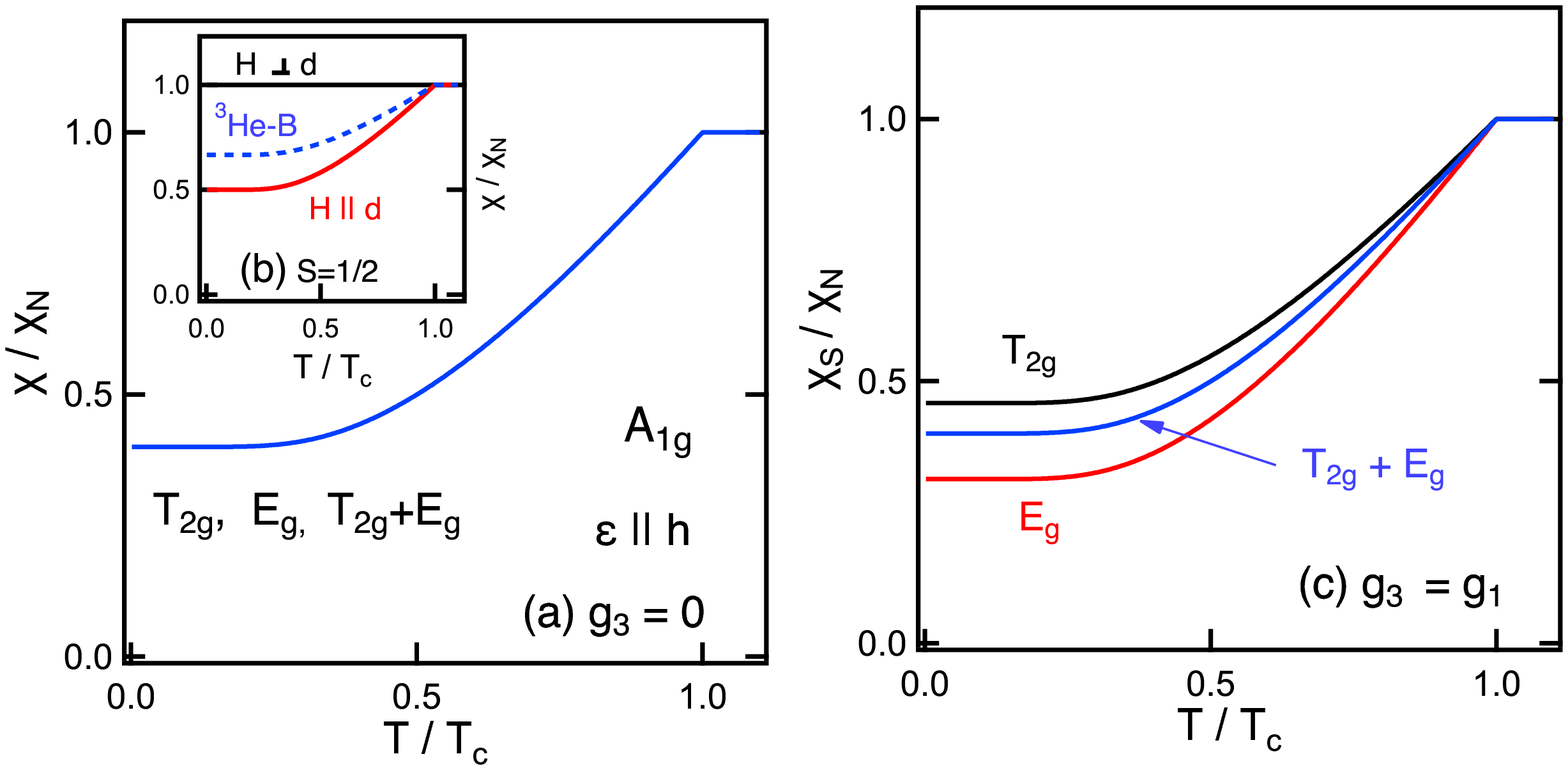}  
\caption{
The spin susceptibility for $A_{1g}$ states in the pseudospin-quintet superconductors is plotted as a function of temperature 
in (a), where we consider $g_3=0$ and $\vec{\epsilon}=\vec{h}$. 
The diagonal elements in the spin susceptibility tensor are isotropic in real space 
and the off-diagonal elements are zero.
In (b), the susceptibility of spin-triplet superconductors are shown for
a helical spin-triplet superconductor with two solid lines and for a $^3$He B-phase 
with a broken line.
In (c), we consider the effects of $J_j^3$ terms in the Zeeman Hamiltonian by choosing
$g_3=g_1$.
Although the amplitudes of susceptibility deviate slightly from those in (a), 
the characteristic features of the susceptibility retain.
}
\label{fig:d-wave}
\end{center}
\end{figure}
%\twocolumngrid
%----------------------------
%

As shown in the second term in Eq.~(\ref{eq:fs_general}),
the anomalous Green's function contains the pairing 
correlation belonging to odd-frequency symmetry class. 
 The stable superconducting states can be described by 
\begin{align}
  [\vec{\eta}_{\boldsymbol{k}} \cdot \vec{\gamma}, 
 \vec{\epsilon}_{\boldsymbol{k}} \cdot \vec{\gamma}]_{-}=0,
 \label{eq:odd_less}
\end{align}
which means the absence of odd-frequency pairs.
Odd-frequency pairs increase the free-energy of a uniform superconducting 
 state~\cite{asano:prb2015} because they indicate the paramagnetic response to 
 a magnetic field~\cite{tanaka:prb2005,asano:prl2011,suzuki:prb2014}.
A phenomenological argument on the paramagnetic response of odd-frequency Cooper pairs 
is given in Appendix A of Ref.~\cite{shu:prb2022}.
 Eq.~(\ref{eq:odd_less}) gives a guide that relates the stable pair potential $\vec{\eta}$ 
to the electronic structures $\vec{\epsilon}$. 
To understand this situation, we briefly summarize the case of spin-triplet superconductors  
in the presence of a strong Rashba spin-orbit interaction $\boldsymbol{\lambda} \cdot \boldsymbol{\sigma}$ 
with 
\begin{align}
\boldsymbol{\lambda} = \lambda_{\mathrm{so}}( \hat{k}_y \boldsymbol{e}_x - \hat{k}_x \boldsymbol{e}_y), \label{eq:rashba}
\end{align}
where $\lambda_{\mathrm{so}}$ represents the amplitude of spin-orbit interaction, 
$\sigma_j$ and $\boldsymbol{e}_j$
  for $j=x, y$ and $z$ are the Pauli matrix and the unit vector in spin space, respectively.
The stable order parameter $ i \boldsymbol{d} \cdot \boldsymbol{\sigma} \, \sigma_y$ is determined as
\begin{align}
\left[ \boldsymbol{d} \cdot \boldsymbol{\sigma}, \boldsymbol{\lambda} \cdot \boldsymbol{\sigma} \right]_{-}=0
\quad  \text{or} \quad \boldsymbol{d} \parallel \boldsymbol{\lambda}, \label{eq:frigeri}
\end{align}
 so that odd-frequency pairs are absent and 
the transition temperature is optimal~\cite{asano:prb2015,frigeri:njp2004}. 
Namely, the order parameter of a spin helical state is stable in this case.
The pair potentials other than the helical state would be realized when the 
Rashba spin-orbit interaction is sufficiently weak.
%The odd-frequency pairing correlations in a spin-triplet superconductor 
%is proportional to $\omega_n \boldsymbol{\lambda} \times \boldsymbol{d}$.  
The choice in Eq.~(\ref{eq:frigeri}) and that in Eq.~(\ref{eq:odd_less}) are equivalent 
to each other. 
We choose the normal state dispersion $\vec{\epsilon}_{\boldsymbol{k}}$ 
so that Eq.~(\ref{eq:odd_less}) is satisfied for the pair potentials 
in Eqs.~(\ref{eq:eta_d})-(\ref{eq:eta_t2g}).  
The Green's function in the superconducting state can be expressed 
simply and analytically under Eq.~(\ref{eq:odd_less}).  
The results of the Green's function in such a case are shown in Appedix~\ref{ap:super}.

The spin susceptibility results in 
\begin{align}
\frac{\chi_{\mu \nu}}{\chi_{\mathrm{N}}}&= \delta_{\mu, \nu} \nonumber\\
-&  \pi T \sum_{\omega_n} \left\langle \frac{1}{2\Omega^3}
\left( {\vec{\eta}^{ }}^2 \, \delta_{\mu, \nu} + 
\frac{ {L}_{\mu, \nu}(\vec{\eta}) }{P_+} \right) \right\rangle_{\hat{\boldsymbol{k}}},\label{eq:chi1}
\end{align}
with $\Omega=\sqrt{\omega_n^2+{\vec{\eta}^{ }}^2}$. The tensor is defined by
\begin{align}
L_{\mu, \nu}(\vec{\eta}) \equiv & \mathrm{Tr}\left[ 
 \vec{\eta} \cdot \vec{\gamma}\, \tilde{J}_\mu\, \vec{\eta} \cdot \vec{\gamma}\tilde{ J}_\nu
\right], \label{eq:l_def}
\end{align}
and its elements are calculated to be
\begin{align}
{L}_{xx} =& P_- ( \eta_1^2 - \eta_2^2 +\eta_3^2) + R_+ \eta_4^2 -R_- \eta_5^2 \nonumber\\
& - 4\sqrt{3} 
p_1^2 \, \eta_4\, \eta_5,\label{eq:l_xx}\\ 
{L}_{yy} =& P_- ( \eta_1^2 + \eta_2^2 -\eta_3^2) + R_+ \eta_4^2  -R_-
 \eta_5^2 \nonumber\\
&+ 4\sqrt{3} 
p_1^2 \, \eta_4\, \eta_5,\label{eq:l_yy}\\ 
{L}_{zz} =&P_- ( - \eta_1^2 + \eta_2^2 +\eta_3^2 -\eta_4^2 ) + P_+ \eta_5^2,\label{eq:l_zz}\\ 
{L}_{xy} =& 2\left[ P_+ \, \eta_2\, \eta_3 - 2\sqrt{3} \,p_1 p_2 \eta_1 \, \eta_5\right],\label{eq:l_xy}\\ 
{L}_{yz} =& 2\left[ P_+ \, \eta_1\, \eta_3 + \sqrt{3} \,p_1 p_2 (\eta_2 \, \eta_5 - \sqrt{3}\, \eta_2 \eta_4)\right],\\ 
{L}_{zx} =& 2\left[ P_+ \, \eta_1\, \eta_2 + \sqrt{3} \, p_1\, p_2 ( \eta_3 \, \eta_5 + \sqrt{3} \, \eta_3 \eta_4)\right], \label{eq:l_zx}\\
P_\pm \equiv& 4p_1^2 \pm p_2^2, \quad R_\pm = 2p_1^2 \pm p_2^2.
\end{align}
The results in Eq.~(\ref{eq:l_xx})-(\ref{eq:l_zx}) describe the characteristic features of the 
susceptibility of $J=3/2$ superconductors.

% ------------------------------------
\subsection{$A_{1g}$ state}
% ------------------------------------

In the presence of attractive interactions in an $A_{1g}$ channel, 
the pair potential is represented by Eqs.~(\ref{eq:pairpotential0}) and\ref{eq:eta_d}).
%\begin{align}
%\vec{\eta}_{\boldsymbol{k}} \cdot \vec{\gamma}  = 
%\Delta \, \sum_{j=1}^5 h_j\, c_j(\hat{\boldsymbol{k}})\, \gamma^j, \label{eq:eta_d}
%\end{align}
%where the unit vector $\vec{h}$ determines the pseudospin structure of the pair potential.
We always set $\vec{e} = \vec{h}$ so that Eq.~(\ref{eq:odd_less}) is satisfied.
Different from the $s$-wave spin-singlet pairing that also belongs to the $A_{1g}$ irrep, an $A_{1g}$ state in the pseudo-spin quintet pairing has the pseudospin degrees of freedom, which allows us to choose a variety of pseudospin structures. 
We first choose a pseudospin structure of a 
$T_{2g}$ irrep characterized by $\vec{h}=\vec{h}_{T_{2g}}=(1, 1, 1, 0, 0)/\sqrt{3}$. 
The spin susceptibility is calculated as 
\begin{align}
\chi_{\mu \nu}=&\chi_{\mathrm{N}} \,  \delta_{\mu, \nu} 
\left[ 1- \frac{1}{2}\left( 1 + \frac{P_- }{3 P_+ }  \right) \mathcal{Q}\right], \label{eq:d_t2g}\\
\mathcal{Q}(T) =& \pi T \sum_{\omega_n}\frac{{\Delta}^2}{\Omega^3}. 
\end{align}
In BCS theory, $\mathcal{Q}$ represents the fraction of Cooper pairs to quasiparticles 
on the Fermi surface. Indeed $\mathcal{Q}$ is zero at $T=T_c$,  
 increases monotonically with the decrease of $T$, and becomes unity at $T=0$.
 The susceptibility tensor is diagonal and isotropic in real space.  
The susceptibility for a pseudospin structure of an $E_{g}$ irrep characterized by 
$\vec{h}=\vec{h}_{E_{g}}=(0, 0, 0, 1, 1)/\sqrt{2}$ results in,
\begin{align}
\chi_{\mu\nu}=&\chi_{\mathrm{N}} \,  \delta_{\mu, \nu}  
\left[ 1- \frac{1}{2}\left( 1 + \frac{p_2^2}{P_+ }  \right) \mathcal{Q}\right]. 
\label{eq:d_eg}
\end{align}
The susceptibility for an admixture of $T_{2g}$ and $E_g$ irreps (
 $\vec{h}_{T_{2g}+E_g}=(1, 1, 1, 1, 1)/\sqrt{5}$) is also calculated as
\begin{align}
\chi_{\mu\nu}=&\chi_{\mathrm{N}} \, \delta_{\mu, \nu}  
\left[ 1- \frac{1}{2}\left( 1 + \frac{1}{5}  \right) \mathcal{Q}\right]. 
\label{eq:d_t2geg}
\end{align}
The off-diagonal elements in the susceptibility tensor are always zero in these cases, ($\chi_{\mu\nu}=0$ for 
$\mu \neq \nu$).

%
%-----------------------------------------------------
%----------------------------
%\onecolumngrid
\begin{figure*}[t]
\begin{center}
  \includegraphics[width=16.0cm]{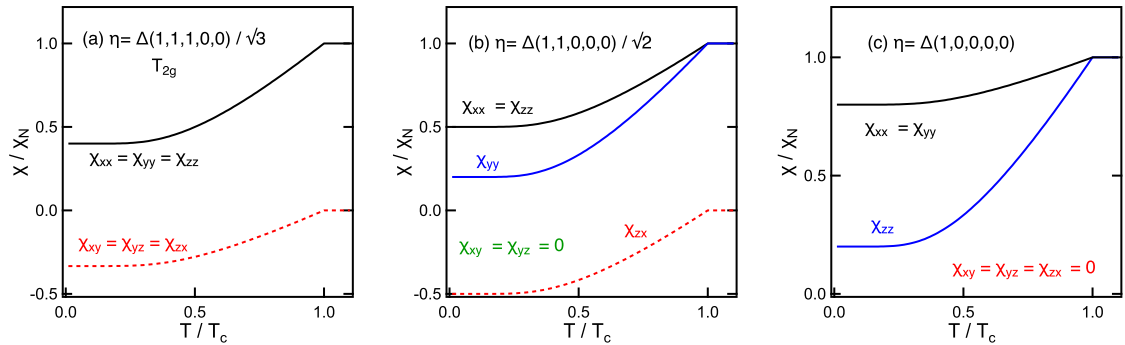}  
\caption{
The spin susceptibility is plotted as a function of temperature
 for a $T_{2g}$ state with $(l_1,l_2,l_3)=(1,1,1)/\sqrt{3}$ in (a).
The susceptibility tensor in a $T_{2g}$ state has off-diagonal elements. 
In (b), the results for the state with $(1,1,0)/\sqrt{2}$ are displayed, where we delete $\gamma^3$ component from the $T_{2g}$ pair potential.
The results for a single component pair potential with $(1,0,0)/\sqrt{2}$ are shown in (c). 
Although we put $g_3=0$ in these figures, $J_j^3$ terms in the Zeeman potential 
do not change the characteristic features.   
}
\label{fig:t2g}
\end{center}
\end{figure*}
%\twocolumngrid
%----------------------------

In Fig.~\ref{fig:d-wave}(a), we plot the susceptibility at $g_3=0$ as a function of temperature.
The dependence of the pair potential on temperature is calculated by solving 
the gap equation in the weak coupling limit. 
As shown in Appedix~\ref{ap:interaction}, the gap equation is common for all the 
superconducting states considered in the present paper and is identical to that in BCS theory.
The results show that the susceptibility 
decreases with the decrease of temperature below $T_c$. 
The results are independent of such choices of the pseudospin structures
as $T_{2g}$, $E_{g}$, and $T_{2g}+E_g$ at $g_3=0$. 
Although a Cooper pair has the angular momentum of $J=2$ in the quintet states, 
the susceptibility is isotropic for all the pseudospin structures. 
The characteristic behaviors are independent of the strength of spin-orbit interaction
 $\beta/\alpha$.
The isotropic feature of the spin susceptibility is considered as a result of high symmetry 
of the pair potential.
The pseudospin structures in the pair potential are characterized by $\gamma^j$ which can be 
described by the linear combination of   
$J_\mu J_\nu$ as shown in Eqs.~(\ref{eq:gamma1})-(\ref{eq:gamma5}). 
For a $T_{2g}$ irrep, $\gamma^j$ for $j=1-3$ characterize both the normal state dispersion and the pair potential.
The pair potential with $\vec{h}_{T_{2g}}$ is symmetric under the cyclic permutation among
$J_x, J_y$ and $J_z$.
For an $E_{g}$ irrep, the pair potential is symmetric under interchanging $J_x \leftrightarrow  J_y$.
The structure of $\vec{h}_{E_{g}}$ results in the isotropic susceptibility including the $z$ direction. 
Thus, the direction of a Zeeman field $\boldsymbol{H}$ in real space does not point
a specific direction in pseudospin space. 
For comparison, we briefly mention the spin susceptibility in 
superfluid $^3$He B-phase described by
\begin{align}
\boldsymbol{d} = \Delta( \hat{k}_x \boldsymbol{e}_x + \hat{k}_y \boldsymbol{e}_y + \hat{k}_z \boldsymbol{e}_z).
\end{align}
The pair potential is symmetric under the cyclic permutation among $x, y$ and $z$.
As a result, the susceptibility plotted with a broken line in Fig.~\ref{fig:d-wave}(b) is isotropic in real space. 

At the end of the subsection, 
we briefly discuss the effects of $J_\mu^3$ term on the spin susceptibility by choosing $g_3=g_1$.
The results are shown in Fig.~\ref{fig:d-wave}(c). 
The isotropic nature of the susceptibility remains unchanged even for $g_3 =g_1$. 
The amplitude of the susceptibility depends on the pseudospin structure of the pair potential; the amplitude for 
a $T_{2g}$ irrep becomes slightly larger than that for an $E_{g}$ irrep.

% ------------------------------------
\subsection{ $T_{2g}$ state}
% ------------------------------------
When attractive interactions between two electrons work in a $T_{2g}$ or an $E_g$ channel, 
the order parameters in Eqs.~(\ref{eq:eta_eg}) and (\ref{eq:eta_t2g})
are isotropic in momentum space. 
To satisfy Eq.~(\ref{eq:odd_less}), we switch off $\vec{\epsilon}=0$ and consider 
a simple pseudospin quintet superconductor in the following subsections.
In other words, the pair potentials independent of momenta
are stable when $|\vec{\epsilon}|$ is sufficiently smaller than the Fermi energy $\mu$.
Even if we choose $\vec{\epsilon}=0$, superconductors show the rich magnetic response 
depending on the pseudospin structures of the pair potential. 
The effects of the pseudospin-dependent dispersions $\vec{\epsilon}\neq 0$ 
on the magnetic response will be discussed later.

%For a $T_{2g}$ state [Eq.~(\ref{eq:eta_t2g})],
We first discuss a $T_{2g}$ state with $(l_1,l_2,l_3)=(1,1,1)/\sqrt{3}$. 
In addition to the diagonal element 
given by Eq.(\ref{eq:d_t2g}), the susceptibility has the off-diagonal elements as
\begin{align}
\chi_{xy} =& \chi_{yz} = \chi_{zx}=
 -\chi_{\mathrm{N}}
\frac{1}{3} \mathcal{Q}.
\end{align}
The results for $g_3=0$ are displayed in Fig.~\ref{fig:t2g}(a). 
The pair potential includes off-diagonal terms $J_\mu\, J_\nu$ with $\mu \neq \nu$ through 
$\gamma^j$ for $j=1-3$ as shown in Eqs.~(\ref{eq:gamma1})-(\ref{eq:gamma5}).
The first term of $L_{\mu\nu}(\vec{\eta})$ in Eqs.~(\ref{eq:l_xy})-(\ref{eq:l_zx}) is the direct results 
of such pseudospin structure.
Since the pair potential is independent of wavenumber, these off-diagonal terms 
remain nonzero values even after averaging over directions in momentum space on the Fermi surface.
Thus, the appearance of the off-diagonal elements in the susceptibility tensor 
is a characteristic feature of a $T_{2g}$ state.
 
Secondly, we display the susceptibility for a $T_{2g}$ state with $(l_1,l_2,l_3)=(1,1,0)/\sqrt{2}$ 
 in Fig.~\ref{fig:t2g}(b), where we delete $\gamma^3$ component from Eq.~(\ref{eq:eta_t2g}).
As a result, $J_y$ is no longer equivalent to $J_x$ and $J_z$, which explains 
the anisotropy of the diagonal elements in Fig.~\ref{fig:t2g}(b).
As shown in Eqs.~(\ref{eq:l_xy})-(\ref{eq:l_zx}), the off-diagonal elements are finite
only for the multi-component pair potentials. At the present case, only the $\chi_{zx}$ 
element remains finite because of $l_3=0$.

Finally, we display the susceptibility for a $T_{2g}$ state with $(l_1,l_2,l_3)=(1,0,0)/\sqrt{2}$ in Fig.~\ref{fig:t2g}(c), which has only $\gamma^1$ component. 
The diagonal elements are anisotropic as they are in Fig.~\ref{fig:t2g}(b). 
All the off-diagonal elements vanish because the pair potential has only one pseudospin component. 
Thus, the off-diagonal elements emerge if $\gamma^i$ and $\gamma^j$ ($i,j=1,2,3; \, i\neq j $) coexist in the pair potential. In particular, the nonzero off-diagonal elements are determined from the direction of high symmetry axis in tetragonal symmetric superconductors, e.g., $\chi_{xy}=\chi_{yz}=0$ and $\chi_{zx} \neq 0$ for the pair potential with $(l_1,l_2,l_3)=(1,1,0)/\sqrt{2}$ since the high symmetry axis is the $y$ direction.

% ------------------------------------
\subsection{ $E_{g}$ state}
% ------------------------------------

\begin{figure}[t]
\begin{center}
  \includegraphics[width=8.0cm]{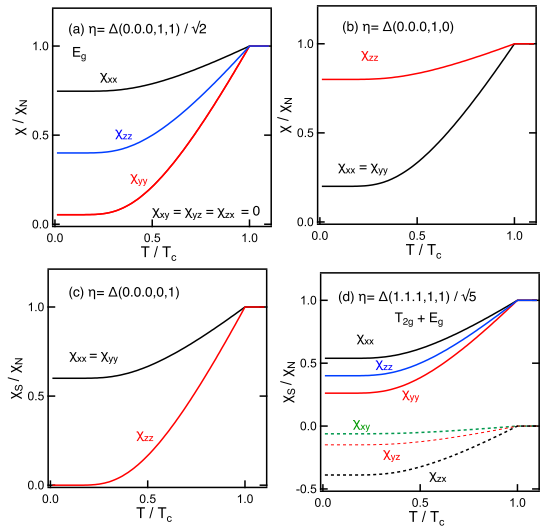}  
\caption{
The spin susceptibility is plotted as a function of temperature
 for a $E_{g}$ state in (a).
}
\label{fig:eg}
\end{center}
\end{figure}

The susceptibility for an $E_{g}$ state with $(l_4,l_5)=(1,1)/\sqrt{2}$ is calculated as
\begin{align}
\chi_{xx} =& 
\chi_{\mathrm{N}}
\left[ 1 - \frac{1}{2}\left( 1 + \frac{p_2^2 - \sqrt{3}p_1^2}{P_+}\right) \mathcal{Q} \right], \\
\chi_{yy}=&\chi_{\mathrm{N}}
\left[ 1 - \frac{1}{2}\left( 1 + \frac{p_2^2 + \sqrt{3}p_1^2}{P_+}\right)  \mathcal{Q} \right], \\
 \chi_{zz}=& \chi_{\mathrm{N}}
\left[ 1 - \frac{1}{2}\left( 1 + \frac{p_2^2}{P_+} \right)  \mathcal{Q} \right], \\
\chi_{xy} =& \chi_{yz} = \chi_{zx}
= 0.
\end{align}
The results are shown in Fig.~\ref{fig:eg}(a). 
All the off-diagonal elements vanish as shown in Eqs.~(\ref{eq:l_xy})-(\ref{eq:l_zx}).
The product of $\gamma^4\, \gamma^5$ does not include such off-diagonal 
terms as $J_\mu J_\nu$ with $\mu \neq \nu$, which explains the absence of the off-diagonal 
elements in the susceptibility tensor.
The diagonal elements of the susceptibility becomes anisotropic due to the last term in Eqs.~(\ref{eq:l_xx}) and (\ref{eq:l_yy}).
The product of $\gamma^4\, \gamma^5$ includes $-J_x^4 + J_y^4$, which
breaks symmetry between $J_x$ and $J_y$.
As a result, the last term in Eq.~(\ref{eq:l_xx}) and that in Eq.~(\ref{eq:l_yy}) have the opposite signs to each other.
Thus the anisotropy in the susceptibility 
is a characteristic feature of an $E_{g}$ state.
The degree of the anisotropy in an $E_g$ state is rather weaker than that in a spin-triplet superconductor.
For comparison, in Fig.~\ref{fig:d-wave}(b), we plot the susceptibility of a spin-triplet 
helical state characterized by
$\boldsymbol{d} \parallel \boldsymbol{\lambda}$ in Eq.~(\ref{eq:rashba}) with two solid lines.
When a Zeeman field is perpendicular to $\boldsymbol{d}$, the susceptibility is a constant across $T_c$. 
The results for $\boldsymbol{H} \parallel \boldsymbol{d}$, on the other hand, the susceptibility 
decreases down to $(1/2) \chi_{\mathrm{N}}$. 
These behaviors are independent of the amplitudes of $\lambda_{\mathrm{so}}>0$. 
In spin-triplet superconductors, both the dimension in spin space 
and that in real space are three. 
Therefore, it is possible to define two different
directions relatively to the direction of a Zeeman field: ($\boldsymbol{d} \parallel \boldsymbol{H}$ 
 and $\boldsymbol{d} \perp \boldsymbol{H}$).
 The clear anisotropy of the susceptibility in Fig.~\ref{fig:d-wave}(b) 
 is a result of the dimensional consistency between in spin space and in real space.
 
In Fig.~\ref{fig:eg}(b) and (c), we display
the results for
the single component states with $(l_4,l_5)=(1,0)$ and $(0,1)$, respectively.
They are possible order parameters of $E_g$ states in the presence of time-reversal symmetry~\cite{sigrist:rmp1991}.
We find the relation
\begin{align}
\chi_{xx}=\chi_{yy} \neq \chi_{zz}, \label{eq:aniso_single}
\end{align}
because the last term in Eqs.~(\ref{eq:l_xx}) and (\ref{eq:l_yy}) is absent.
Including the results in Fig.~\ref{fig:t2g}(c), Eqs.~(\ref{eq:l_xx})-(\ref{eq:l_zx}) suggest that 
the anisotropic response like Eq.~(\ref{eq:aniso_single}) 
and the absense of off-diagonal elements are the common feature of the single component 
order parameter.

%
%-----------------------------------------------------
%----------------------------
%\onecolumngrid
%\twocolumngrid
%----------------------------

% ------------------------------------
\subsection{ Admixture of $T_{2g}$ and $E_g$ states}
% ------------------------------------

The results for a the admixture of $T_{2g}$ and $E_g$ states, (i. e.,  $(l_1,l_2,l_3,l_4,l_5) =(1,1,1,1,1)/\sqrt{5}$), 
 are calculated as
\begin{align}
\chi_{xx} =& 
\chi_{\mathrm{N}}
\left[ 1 - \frac{1}{2}\left( 1 + \frac{P_+ - 4\sqrt{3} p_1^2}{5 P_+} \right) \mathcal{Q} \right], 
\\
\chi_{yy}=&\chi_{\mathrm{N}}
\left[ 1 - \frac{1}{2}\left( 1 + \frac{P_+ + 4\sqrt{3} p_1^2}{5 P_+} \right) \mathcal{Q} \right], \\
 \chi_{zz}=&
\chi_{\mathrm{N}}
\left[ 1 - \frac{1}{2}\left( 1 + \frac{1}{5} \right) \mathcal{Q} \right].
\end{align}
The off-diagonal elements are calculated in the similar way,
\begin{align}
\chi_{xy} =& 
 -\chi_{\mathrm{N}} \, \mathcal{Q}\, \frac{P_+ - 2\sqrt{3}\, p_1\, p_2}{5P_+} , 
\\
\chi_{yz} = &
 -\chi_{\mathrm{N}} \mathcal{Q}\, \frac{P_+ + \sqrt{3}\, p_1\, p_2 (1-\sqrt{3})}{5P_+}, \\
\chi_{zx} =& 
 -\chi_{\mathrm{N}} \mathcal{Q}\, \frac{P_+ + \sqrt{3}\, p_1\, p_2 (1+\sqrt{3})}{5P_+}.
\end{align}
The calculated results for $g_3=0$ are plotted in Fig.~\ref{fig:eg}(d). 
Not only the diagonal elements but also the off-diagonal 
elements are anisotropic. 
All the elements in Eqs.~(\ref{eq:l_xx})-(\ref{eq:l_zx}) 
are finite and different from one another.
The degree of the anisotropy in the diagonal elements are 
weaker than that in $E_g$ state and stronger than that in $T_{2g}$ state.
The characteristic features of the susceptibility displayed in Figs.~\ref{fig:t2g} and \ref{fig:eg} 
retain even if we consider $J_\mu^3$ term in the Zeeman Hamiltonian.

Finally, we briefly discuss the effects of the pseudospin-dependent dispersion $\vec{\epsilon}$ 
on the characteristic behaviors of the susceptibility. 
When we switch on $\vec{\epsilon}$ in $T_{2g}$ and $E_{g}$ states, 
Eq.~(\ref{eq:odd_less}) is no longer holds.
As a result, 
 the additional terms such as
\begin{align}
\frac{C_f}{2} \, i\omega_n\,  [\vec{\eta}_{\boldsymbol{k}} \cdot \vec{\gamma}, \vec{\epsilon}_{\boldsymbol{k}}
\cdot \vec{\gamma}]_{-}
= C_f \, i \omega_n \, \sum_{i \neq j} \epsilon_i \, \eta_j \,  \gamma^i\, \gamma^j,
\end{align}
appear at the numerator of the anomalous Green's function in Eq.~(\ref{eq:f_gene}), 
where $C_f$ is a constant.  
Such components represent the admixture of pseudospin-triplet and the 
pseudospin-septet pairing correlations~\cite{kim:jpsj2021}. 
Their contribution to the susceptibility tensor is proportional to
\begin{align}
C_f^2 \, \omega_n^2 \, \mathrm{Tr}\left[ 
\sum_{i \neq j} \epsilon_i\, \eta_j \gamma^i \gamma^j \; \tilde{J}_\mu\;  
\sum_{k \neq l} \epsilon_k\, \eta_l \gamma^k \gamma^l \; \tilde{J}_\nu
\right], 
\end{align} 
which modify the susceptibility tensor, 
However, they do not always cancel the off-diagonal elements in Fig.~\ref{fig:t2g}(a) in a $T_{2g}$ state. 
They do not wash out the anisotropy of the diagonal elements in Fig.~\ref{fig:eg} in $E_g$ states.

The results in Figs.~\ref{fig:d-wave}-\ref{fig:eg} suggest that the behaviors 
of the susceptibility depends sensitively on the orbital symmetry and the 
pseudospin structures of the pair potential. In particular, the appearance of 
the off-diagonal elements in the susceptibility tensor is a unique feature to $J=3/2$ superconductors.

%%%%%%%%%%%%%%%%%%%%%%%%%%%%%%%%%%%%%%%%%%%%
\section{Conclusion}
%%%%%%%%%%%%%%%%%%%%%%%%%%%%%%%%%%%%%%%%%%%%

We have studied theoretically the spin susceptibility of pseudospin-quintet 
pairing states in a $J=3/2$ superconductor that preserves cubic lattice symmetry and time-reversal symmetry.
Within the linear response to a Zeeman field, we calculate the spin susceptibility by using the Green's 
function that is obtained by solving the Gor'kov equation analytically. 
The pair potentials are chosen so that a superconducting state is stable under the pseudospin structures in the 
normal state.
The calculated results indicat that the magnetic response of pseudospin-quintet states depends sensitively
on the pseudospin structures of the pair potential.
The susceptibility tensor in $A_{1g}$ states is isotropic in real space as that 
in the B-phase of superfluid $^3$He. 
For $E_{g}$ states, the susceptibility tensor becomes anisotropic in real space. 
We found in a $T_{2g}$ state that the susceptibility tensor has the off-diagonal elements.

%---------------------------------------
\begin{acknowledgments}
The authors are grateful to R.~Nomura for useful discussion.
This work was supported by JSPS KAKENHI (Nos. JP19K14612, JP20H01857, JP22K03478) and 
JSPS Core-to-Core Program (No. JPJSCCA20170002). 
T.~S. is supported in part by the establishment of university 
fellowships towards the creation of science technology innovation from the 
Ministry of Education, Culture, Sports, Science, and Technology (MEXT) of Japan. 
S.K. was supported by the CREST project (Grants No. JPMJCR19T2) from Japan Science and Technology Agency (JST).
\end{acknowledgments}
%---------------------------------------

\appendix
\begin{widetext}
% =========================================
\section{Normal state}\label{ap:normal}
% =========================================

The spin susceptibility in the normal state is give by
\begin{align}
\chi_{\mathrm{N}}=& -
\left(\frac{\mu_B}{2}\right)^2 
 T\sum_{\omega_n} 
\int\frac{d \boldsymbol{k}}{(2\pi)^d} 
 \mathrm{Tr} \left[
{G}_{\mathrm{N}}(\boldsymbol{k}, i\omega_n) \, J_\mu 
\,
G_{\mathrm{N}}(\boldsymbol{k}, i\omega_n) \, J_\nu
\right], \\
G_{\mathrm{N}} =& \frac{i\omega_n - \xi + \vec{\epsilon}_{\boldsymbol{k}} \cdot \vec{\gamma} }
{(i\omega_n-\xi)^2 - {\vec{\epsilon}_{\boldsymbol{k}}}^2}
= \alpha_N + \beta_N \vec{\epsilon}_{\boldsymbol{k}} \cdot \vec{\gamma},\quad
\alpha_N= \frac{1}{2}
\left[\frac{1}{z_{N+}} + \frac{1}{z_{N-}} \right], \quad 
\beta_N=
\frac{1}{2 |\vec{\epsilon_{\boldsymbol{k}}}|}
\left[ \frac{1}{z_{N+}} - \frac{1}{z_{N-}}\right], \\
z_{N\pm}=&i\omega_n - \xi_\pm, \quad \xi_\pm = \xi \pm |\vec{\epsilon_{\boldsymbol{k}}}|.
\end{align}
The the Green's function is the solution of
\begin{align}
\left[ i\omega_n - H_{\mathrm{N}} \right]G_{\mathrm{N}}(\boldsymbol{k}, \omega_n)=1, \quad 
H_{\mathrm{N}}(\boldsymbol{k}) = \xi_{\boldsymbol{k}} + \vec{\epsilon}_{\boldsymbol{k}} \cdot \vec{\gamma}.
\end{align}
The trace of the Green's function is calculated as
\begin{align}
\mathrm{Tr} \left[
{G}_{\mathrm{N}}(\boldsymbol{k}, i\omega_n) \, \tilde{J}_\mu 
\,
G_{\mathrm{N}}(\boldsymbol{k}, i\omega_n) \, \tilde{J}_\nu
\right]
=  \delta_{\mu, \nu} \, P_+ \, \alpha_N^2 + \alpha_N\, \beta_N M_{\mu, \nu}(\vec{\epsilon}_{\boldsymbol{k}})
+ \beta_N^2 L_{\mu, \nu}(\vec{\epsilon}_{\boldsymbol{k}}),
\end{align}
where we use $\mathrm{Tr}( \tilde{J}_\mu \tilde{J}_\nu) = P_+ \, \delta_{\mu, \nu}$ and define the tensor
\begin{align}
M_{\mu, \nu}(\vec{\epsilon} )\equiv  \mathrm{Tr}\left[ \vec{\epsilon} \cdot \vec{\gamma} 
(\tilde{J}_\mu\, \tilde{J}_\nu + \tilde{J}_\nu \tilde{J}_\mu)\right]. 
\end{align}
The angular momenta $J_\nu$ are expressed in terms of $\gamma^\nu$ 
\begin{align}
J_x=& \frac{-i}{2}\left[\sqrt{3} \gamma^2\, \gamma^5 +\gamma^1\, \gamma^3 + \gamma^2\, \gamma^4 \right], \quad
J_y= \frac{i}{2}\left[\sqrt{3} \gamma^3\, \gamma^5 +\gamma^1\, \gamma^2 - \gamma^3\, \gamma^4 \right], \quad
J_z= \frac{i}{2}\left[ \gamma^2\, \gamma^3 +2 \, \gamma^1\, \gamma^4 \right].\label{eq:j_gamma}
\end{align}
They obey the relation $U_T\, (J^\nu)^\ast U_T^{-1} = - J^\nu$, which simply means that the angular momenta are antisymmetric under the time-reversal operation. 
The expression of $J_\nu^3$ 
\begin{align}
J_x^3 =& -\frac{i}{8}\left( 7\sqrt{3} \gamma^2\, \gamma^5 + 13 \gamma^1\, \gamma^3 + 7 \gamma^2\, \gamma^4 \right), \;
J_y^3 = \frac{i}{8}\left( 7\sqrt{3} \gamma^3\, \gamma^5 + 13 \gamma^1\, \gamma^2 - 7 \gamma^3\, \gamma^4 
\right),\\
J_z^3 =&  \frac{i}{8} \left[ 13\gamma^2\, \gamma^3  + 14\gamma^1 \, \gamma^4 \right],
\end{align}
suggests that $J_\nu$ and $J_\nu^3$ share the common matrix structures.
The elements of the tensor are calculated as
\begin{align}
M_{xx}(\vec{\epsilon})=& 4 \, p_1\, p_2, (\sqrt{3} \epsilon_4 - \epsilon_5), \quad 
M_{yy}(\vec{\epsilon})= - 4 \, p_1\, p_2,  (\sqrt{3} \epsilon_4 + \epsilon_5), \quad
M_{zz}(\vec{\epsilon})=  8  \, p_1\, p_2,  \epsilon_5, \label{eq:m_res1}\\
M_{xy}(\vec{\epsilon})=M_{yx}(\vec{\epsilon})=& 4\sqrt{3} p_1^2 \epsilon_1, \quad 
M_{xz}(\vec{\epsilon})=M_{zx}(\vec{\epsilon})= 4\sqrt{3} p_1^2 \epsilon_3, 
\quad M_{yz}(\vec{\epsilon})=M_{zy}(\vec{\epsilon})= 4\sqrt{3} p_1^2 \epsilon_2, \label{eq:m_res2}
\end{align}
where $\epsilon_j = \beta \boldsymbol{k}^2 e_j\, c_j(\hat{\boldsymbol{k}}) $ as defined in 
Eq.~(\ref{eq:ep_def}) and we have used the relations
\begin{align}
\mathrm{Tr}[ \gamma^j]=&0, \quad \mathrm{Tr}[ \gamma^i\, \gamma^j]=4 \delta_{i,j}, \quad
\mathrm{Tr}[ \gamma^i\, \gamma^j\, \gamma^k ]=0, \quad
\mathrm{Tr}[ \gamma^i\, \gamma^j\, \gamma^k \, \gamma^l]=
4 \left[\delta_{i,j}\delta_{k,l}- \delta_{i,k}\delta_{j,l} + \delta_{i,l}\delta_{j,k} \right].
\end{align}
The another tensor $L_{\mu, \nu}$ is defined by Eq.~(\ref{eq:l_def}).

The summation over the wavenumber is replaced by the integration as
\begin{align}
\int\frac{d \boldsymbol{k}}{(2\pi)^d} F(\boldsymbol{k}) \to N_0 \int_{-\infty}^\infty d \xi\, 
\langle F(\xi, \hat{\boldsymbol{k}}) \rangle_{\hat{\boldsymbol{k}}},\quad
\langle F(\xi, \hat{\boldsymbol{k}}) \rangle_{\hat{\boldsymbol{k}}}= \int\frac{d \hat{\boldsymbol{k}}}{4\pi} 
 F(\xi, \hat{\boldsymbol{k}}), 
\end{align}
where $\hat{\boldsymbol{k}}$ is the unit vector on the Fermi surface. 
By using the relations 
\begin{align}
\langle M_{\mu, \nu}(\vec{\epsilon}_{\boldsymbol{k}}) \rangle_{\hat{\boldsymbol{k}}}=0, \quad
\langle L_{\mu, \nu}(\vec{\epsilon}_{\boldsymbol{k}}) \rangle_{\hat{\boldsymbol{k}}}= \langle L_{\mu, \mu} \rangle_{\hat{\boldsymbol{k}}} \, \delta_{\mu, \nu}.
\end{align}
the susceptibility in the normal state becomes
\begin{align}
\chi_{\mathrm{N}}=& -
\left(\frac{\mu_B}{2}\right)^2 
 T\sum_{\omega_n} 
\int\frac{d \boldsymbol{k}}{(2\pi)^d} 
 \left[
P_+ \alpha_N^2 \, + \beta_N^2 \langle L_{\mu, \mu} \rangle_{\hat{\boldsymbol{k}}} 
\right] \, \delta_{\mu, \nu}. 
\end{align}
The summation over the Matsubara frequency is carried out as
\begin{align}
T\sum_{\omega_n} \alpha_N^2 =  &
\frac{1}{4} T\sum_{\omega_n} \left[\frac{1}{z_{N+}^2} + \frac{1}{|\vec{\epsilon_{\boldsymbol{k}}}|}
\left( \frac{1}{z_{N+}} - \frac{1}{z_{N-}} \right) + \frac{1}{z_{N-}^2}
\right],\\
=&\frac{1}{4}\left[
-\frac{1}{4T}\cosh^{-2}\frac{\xi_+}{2T} - \frac{1}{2|\vec{\epsilon_{\boldsymbol{k}}}|} 
\left( \tanh \frac{\xi_+}{2T} - \tanh \frac{\xi_-}{2T} \right) 
-\frac{1}{4T}\cosh^{-2}\frac{\xi_-}{2T} 
 \right].
\end{align}
The integration over $\xi$ after the summation over the frequency can be calculated exactly as
\begin{align}
 \int_{-\infty}^\infty d\xi \, T\sum_{\omega_n} \alpha_N^2 = -1, \quad 
\int_{-\infty}^\infty d\xi \, T\sum_{\omega_n} \beta_N^2 =0.
\end{align}
The resulting spin susceptibility
\begin{align}
\chi_{\mathrm{N}}=& \left(\frac{\mu_B}{2}\right)^2 \, P_+ \, N_0,
\end{align}
is diagonal and isotropic independent of the direction of a Zeeman field.

The normal Green's function can be described alternatively as
\begin{align}
G_{\mathrm{N}}=& \frac{-1}{Z_N(\omega_n)}\left[ 
A_N + B_N \vec{\epsilon}_{\boldsymbol{k}} \cdot \vec{\gamma}\right],\quad 
Z_N= \xi^4 + 2\xi^2( \omega_n^2- {\vec{\epsilon}_{\boldsymbol{k}} }^2) 
+ ( \omega_n^2+ {\vec{\epsilon}_{\boldsymbol{k}} }^2)^2\\
A_N=&
(\omega_n^2+ \xi^2 + {\vec{\epsilon}_{\boldsymbol{k}} }^2) i\omega_n 
+  (\omega_n^2+ \xi^2 - {\vec{\epsilon}_{\boldsymbol{k}} }^2) \xi, \quad 
B_N = -\left\{ (i\omega_n - \xi)^2 - {\vec{\epsilon}_{\boldsymbol{k}} }^2 \right\}.
\end{align}
When we carry out the summation over the wavenumber first as
\begin{align}
N_0 \int_{-\infty}^{\infty} d\xi &\langle
\mathrm{Tr} \left[
{G}_{\mathrm{N}}(\boldsymbol{k}, i\omega_n) \, \tilde{J}_\mu 
\,
G_{\mathrm{N}}(\boldsymbol{k}, i\omega_n) \, \tilde{J}_\nu
\right]
 \rangle_{\hat{\boldsymbol{k}}}= N_0 \int_{-\infty}^{\infty} d\xi 
 \frac{1}{Z_N^2}
\left[ \delta_{\mu, \nu}\, P_+ \, A_N^2 + B_N^2 \langle L_{\mu, \nu} \rangle_{\hat{\boldsymbol{k}}} \right],
\end{align}
we find $\chi_{\mathrm{N}}=0$ because of 
\begin{align}
\int_{-\infty}^{\infty} d\xi \,
 \frac{A_N^2}{Z_N^2} = \int_{-\infty}^{\infty} d\xi \,
 \frac{B_N^2}{Z_N^2} =0. \label{eq:anbn_zero}
\end{align}
The discrepancy is derived from the fact that the integration over the wavenumber and the summation 
over the frequency do not converge.\cite{agd}
On the way to Eq.~(\ref{eq:anbn_zero}), 
we have used the following relations
\begin{align}
I_0 =& \int_{-\infty}^\infty d\xi \,\frac{1}{Z_N} = 
\frac{\pi}{2|\omega_n|(\omega_n^2+\varepsilon^2) }, \quad 
J_n =  \int_{-\infty}^\infty d\xi \, \frac{\xi^n}{Z_N^2},\\
J_0 =& \frac{I_0}{8\omega_n^2}\frac{5\omega_n^2+\varepsilon^2}
{(\omega_n^2+\varepsilon^2)^2}, \quad 
J_2 = \frac{I_0}{8\omega_n^2}, \quad 
J_4 =\frac{I_0}{8\omega_n^2}(\omega_n^2+\varepsilon^2),
\quad J_6 = \frac{I_0}{8\omega_n^2}(\omega_n^2+\varepsilon^2)(5\omega_n^2+\varepsilon^2).
\end{align}
We approximately replace $ {\vec{\epsilon}_{\hat{\boldsymbol{k}}}}^2 >0 $ 
by $\varepsilon^2 =(\alpha/\beta +5/4)^{-2} \mu^2$.

% =========================================
\section{Superconducting state}\label{ap:super}
% =========================================
The Green's function under Eq.~(\ref{eq:odd_less}) is calculated to be
\begin{align}
G(\boldsymbol{k}, \omega_n) = &
\frac{1}{Z_{\mathrm{S}}}\left[ A_g + {B_g}
\vec{\epsilon}_{\boldsymbol{k}} \cdot {\vec{\gamma}} \right],\\
A_g=& -(\omega_n^2+ {\xi_{\boldsymbol{k}} }^2 \
+{\vec{\eta}_{\boldsymbol{k}} }^2+ {\vec{\epsilon}_{\boldsymbol{k}} }^2) i\omega_n 
- (\omega_n^2+ {\xi_{\boldsymbol{k}} }^2 +{\vec{\eta}_{\boldsymbol{k}} }
^2 - {\vec{\epsilon}_{\boldsymbol{k}} }^2) \, 
\xi_{\boldsymbol{k}},
\quad B_g= \omega_n^2- \xi^2 - 2 \,i \,\omega_n \, \xi_{\boldsymbol{k}}  
+{\vec{\eta}_{\boldsymbol{k}} }^2+ {\vec{\epsilon}_{\boldsymbol{k}} }^2
,\\
{F}(\boldsymbol{k}, \omega_n)  =&\frac{1}{Z_{\mathrm{S}}}\left[ A_f + {B_f}
 \vec{\eta}_{\boldsymbol{k}}  \cdot {\vec{\gamma}} \right]U_T, \quad 
\undertilde{F}(\boldsymbol{k}, \omega_n) =\frac{U_T}{Z_{\mathrm{S}}}\left[ A_f + {B_f}
\vec{\eta}_{\boldsymbol{k}}  \cdot {\vec{\gamma}} \right], \label{eq:f_gene}\\
A_f=& - 2 \xi_{\boldsymbol{k}}  \, \vec{\epsilon}_{\boldsymbol{k}}  \cdot \vec{\eta}_{\boldsymbol{k}}, \quad
B_f= (\omega_n^2+ {\xi_{\boldsymbol{k}} }^2 +{\vec{\eta}_{\boldsymbol{k}} }^2+ 
{\vec{\epsilon}_{\boldsymbol{k}} }^2),\quad
Z_{\mathrm{S}}= (\omega_n^2+ {\xi_{\boldsymbol{k}} }^2 +{\vec{\eta}_{\boldsymbol{k}} }^2 
+ {\vec{\epsilon}_{\boldsymbol{k}} }^2 )^2 - 4 
{\xi_{\boldsymbol{k}}}^2 {\vec{\epsilon}_{\boldsymbol{k}} }^2
=Z_N(\Omega),
\end{align}
with $\Omega=\sqrt{\omega_n^2 + {\vec{\eta}_{\boldsymbol{k}}}^2}$.
When we carry out the summation over the wavenumber, we find 
\begin{align}
N_0 &\int_{-\infty}^{\infty} d\xi \langle
\mathrm{Tr} \left[
{G}_{\mathrm{S}}(\boldsymbol{k}, i\omega_n) \, \tilde{J}^\mu 
\,
G_{\mathrm{S}}(\boldsymbol{k}, i\omega_n) \, \tilde{J}^\nu
\right] \rangle_{\hat{\boldsymbol{k}}},\nonumber\\
=& N_0 \int_{-\infty}^{\infty} d\xi \left\langle \left[
\left\{\frac{ A_N^2(\Omega)}{Z_N^2(\Omega)} 
+ \frac{{\vec{\eta}_{\boldsymbol{k}}}^2(\Omega^2+ {\vec{\epsilon}_{\boldsymbol{k}} }^2
+\xi^2)^2}{Z_{\mathrm{S}}^2}\right\}\tilde{J}_\mu\, \tilde{J}_\nu
+ \left\{ \frac{ B_N^2(\Omega)}{Z_N^2(\Omega)}
+ \frac{4 \,{\vec{\epsilon}_{\boldsymbol{k}} }^2 \,\xi^2}{Z_{\mathrm{S}}^2}
\right\}   L_{\mu, \nu}(\vec{\eta}_{\boldsymbol{k}}) 
\right] \right\rangle_{\hat{\boldsymbol{k}}},\nonumber\\
=& \left\langle \frac{\pi \, N_0 }{4 \Omega^3(\Omega^2+ \varepsilon^2)}
\left[ P_+ \, {\vec{\eta}_{\boldsymbol{k}}}^2\, (2 \Omega^2+\varepsilon^2) \delta_{\mu, \nu}  
+ \varepsilon^2 \,  {L}_{\mu, \nu}(\vec{\eta}_{\boldsymbol{k}})  \right]
\right\rangle_{\hat{\boldsymbol{k}}},\\
N_0 &\int_{-\infty}^{\infty} d\xi \langle
\mathrm{Tr} \left[
\undertilde{F}_{\mathrm{S}}(\boldsymbol{k}, i\omega_n) \, \tilde{J}^\mu 
\,
F_{\mathrm{S}}(\boldsymbol{k}, i\omega_n) \, (\tilde{J}^\nu)^\ast
\right] \rangle_{\hat{\boldsymbol{k}}},\nonumber\\
=& N_0 \int_{-\infty}^{\infty} d\xi \left\langle \left[
\left\{ \frac{ 4 {\vec{\eta}_{\boldsymbol{k}}}^2\, {\vec{\epsilon}_{\boldsymbol{k}} }^2\, \xi^2)}
{Z_{\mathrm{S}}^2}\right\} \tilde{J}_\mu\, \tilde{J}_\nu
+ \left\{ 
\frac{ (\xi^2 + \Omega^2 + {\vec{\epsilon}_{\boldsymbol{k}} }^2 )^2}{Z_{\mathrm{S}}^2}
\right\} L_{\mu, \nu}(\vec{\eta}_{\boldsymbol{k}})
\right] \right\rangle_{\hat{\boldsymbol{k}}},\nonumber\\
=& \left\langle \frac{\pi N_0 }{4 \Omega^3(\Omega^2+ \varepsilon^2)}
\left[ P_+ \,  {\vec{\eta}_{\boldsymbol{k}}}^2\, \varepsilon^2 \delta_{\mu, \nu} + 
 (2 \Omega^2+\varepsilon^2) \, {L}_{\mu, \nu}(\vec{\eta}_{\boldsymbol{k}})  
 \right] \right\rangle_{\hat{\boldsymbol{k}}}. 
\end{align}
The average $\langle L_{\mu, \nu} \rangle_{\hat{\boldsymbol{k}}}$ describes 
the anisotropy and the off-diagonal response of the spin susceptibility.
%We approximately replace ${\vec{\eta}_{\boldsymbol{k}}}^2>0$ by $\Delta^2$.	

% =========================================
\section{Gap equation}\label{ap:interaction}
% =========================================
The attractive interactions between two electrons are necessary 
for Cooper pairing. Some bosonic excitation usually mediates the attractive 
interactions. In this Appendix, we assume the attractive interaction phenomenologically
and derive the gap equation for superconducting states discussed in this paper.
The pair potential of the superconducting states is defined by
\begin{align}
\Delta_{\alpha, \beta}(\hat{\boldsymbol{k}}) 
= \frac{1}{V_{\mathrm{vol}}} \sum_{\boldsymbol{k}^\prime} \sum_{\lambda, \tau}
V_{\alpha,\beta; \lambda, \gamma}({\boldsymbol{k}}-{\boldsymbol{k}}^\prime) 
\left\langle 
c_{\boldsymbol{k}^\prime, \lambda}\, c_{-\boldsymbol{k}^\prime, \tau}
\right\rangle = -
\frac{1}{V_{\mathrm{vol}}} \sum_{\boldsymbol{k}^\prime} T\sum_{\omega_n}  \sum_{\lambda, \tau}
V_{\alpha,\beta; \lambda, \tau}({\boldsymbol{k}}-{\boldsymbol{k}}^\prime) \,
F_{\lambda, \tau}(\boldsymbol{k}^\prime, \omega_n), \label{eq:def_delta}
\end{align}
where $\alpha$, $\beta$, $\lambda$, and $\tau$ are the indices of pseudospin 
of an electron.
The attractive interaction $V_{\alpha,\beta; \lambda, \tau}$ works 
on two electrons with $\lambda$ and $\tau$
and generats the pair potential between two electrons with $\alpha$ and $\beta$.
The attractive interaction can be decomposed as
\begin{align}
V_{\alpha,\beta; \lambda, \tau}({\boldsymbol{k}}-{\boldsymbol{k}}^\prime)
=\sum_{\nu=1-5}
g_\nu({\boldsymbol{k}}-{\boldsymbol{k}}^\prime)\, (\gamma_\nu\, U_T)_{\alpha, \beta}\, 
(\gamma_\nu\, U_T)^\ast_{\lambda, \tau}.
\end{align}
For $A_{1g}$ states in Sec.~III~A, we choose
\begin{align}
g_\nu(\boldsymbol{k}-\boldsymbol{k}^\prime) =& \left\{
\begin{array}{ll} g \, c_\nu(\hat{\boldsymbol{k}})
\, c_\nu(\hat{\boldsymbol{k}^\prime}) & \nu=1-3 \\
0 & \nu=4, 5
\end{array}\right. \quad T_{2g}\, \textrm{irreps} \label{eq:int_kernel1}\\
g_\nu(\boldsymbol{k}-\boldsymbol{k}^\prime) =& \left\{
\begin{array}{ll} 
0 & \nu=1-3 \\
g \, c_\nu(\hat{\boldsymbol{k}})
\, c_\nu(\hat{\boldsymbol{k}^\prime}) & \nu=4, 5 
\end{array}\right. \quad E_{g}\, \textrm{irreps}\label{eq:int_kernel2}\\
g_\nu(\boldsymbol{k}-\boldsymbol{k}^\prime) =& 
g \, c_\nu(\hat{\boldsymbol{k}})
\, c_\nu(\hat{\boldsymbol{k}^\prime}), \;  \nu=1- 5 \quad T_{2g}+E_{g}\, \textrm{irreps}.
\label{eq:int_kernel3}
\end{align}
By substituting the anomalous Green's function in Eq.~(\ref{eq:f_gene})  
into Eq.~(\ref{eq:def_delta}), we obtain the gap equation 
\begin{align}
\Delta = T\sum_{\omega_n}
g N_0 \int d\xi \frac{ B_f \, \Delta}{Z_\mathrm{S}}, 
\end{align}
where we have used Eq.~(\ref{eq:orthonoaml_cij}).
After integrating over $\xi$, we obtain
\begin{align}
1=g N_0 T\sum_{\omega_n} \frac{1}{\sqrt{\omega_n^2+\Delta^2}}. \label{eq:gap_bcs}
\end{align}
The results coinsides with the gap equation in BCS theory. 
For $T_{2g}$, $E_g$, and an admixture state of them in Sec.~III~B and C, we replace 
$c_\nu(\hat{\boldsymbol{k}})$ by 1 for all $\nu$ in Eqs.~(\ref{eq:int_kernel1})-(\ref{eq:int_kernel3}).
The gap equation for such states is identical to Eq.~(\ref{eq:gap_bcs}).

\end{widetext}

%\bibliography{list_2022}


\end{document}